\documentclass[a4paper,fleqn]{cas-dc}

\usepackage[numbers]{natbib}
\usepackage[inline]{enumitem}

\usepackage{ifthen}  
\usepackage{algorithm}
\usepackage{algpseudocode}

\algnewcommand\KwTo{\textbf{ to }}
\algnewcommand\KwOr{\textbf{ or }}
\algnewcommand\KwAnd{\textbf{ and }}
\algrenewcommand\alglinenumber[1]{\scriptsize #1:}

\def\tsc#1{\csdef{#1}{\textsc{\lowercase{#1}}\xspace}}
\tsc{WGM}
\tsc{QE}
\tsc{EP}
\tsc{PMS}
\tsc{BEC}
\tsc{DE}
\newtheorem{theorem}{Theorem}

\newtheorem{definition}[theorem]{Definition}
\newdefinition{rmk}{Remark}
\newproof{pf}{Proof}
\newproof{pot}{Proof of Theorem \ref{thm2}}

\begin{document}
\let\WriteBookmarks\relax
\def\floatpagepagefraction{1}
\def\textpagefraction{.001}

\shorttitle{Fuzzychain: An Equitable Consensus Mechanism for Blockchain Networks}

\shortauthors{Author}

\title [mode = title]{Fuzzychain: An Equitable Consensus Mechanism for Blockchain Networks}                      



%
\author[1]{Bruno Ramos-Cruz}[type=editor]

\cormark[1]


\ead{brcruz@ujaen.es}



\affiliation[1]{organization={Computer Science Department, University of Jaen},
    addressline={Jaen}, 
    city={Jaen},
    postcode={23071}, 
    country={Spain}}

\author[1,2]{Javier Andreu-Pérez}[]
\ead{j.andreu-perez@essex.ac.uk}

\author[1]{Francisco J. Quesada}[]
\ead{fqreal@ujaen.es}


\affiliation[2]{organization={Centre for Computational Intelligencer, School of Computer Science and Electronic
Engineering, University of Essex},
    city={Colchester},
    country={United Kingdom}}

\author%
[1]
{Luis Martínez}
\ead{martin@ujaen.es}


\cortext[cor1]{Corresponding author}



\begin{abstract}
Blockchain technology has become a trusted method for establishing secure and transparent transactions through a distributed, encrypted network. The operation of blockchain is governed by consensus algorithms, among which Proof of Stake (PoS) is popular yet has its drawbacks, notably the potential for centralising power in nodes with larger stakes or higher rewards. Fuzzychain, our proposed solution, introduces the use of fuzzy sets to define stake semantics, promoting decentralised and distributed processing control. This system selects validators based on their degree of membership to the stake fuzzy sets rather than just the size of their stakes. As a pioneer proposal in applying fuzzy sets to blockchain, Fuzzychain aims to rectify PoS's limitations. Our results indicate that Fuzzychain not only matches PoS in functionality but also ensures a fairer distribution of stakes among validators, leading to more inclusive validator selection and a better-distributed network.
\end{abstract}





\maketitle

\section{Introduction}

The transition to digital business models highlights a key challenge: establishing trust among stakeholders in a virtual environment. Several strategies, including using trusted third parties, digital signatures, distributed systems, and peer-to-peer networks \cite{Sankar2017}, have been explored to address this issue. However, these methods have limitations, steering focus towards blockchain technology as a promising solution to build and maintain trust in digital transactions.

Blockchain, a decentralised, secure, and peer-to-peer network, addresses the challenges of trust and secure transactions in digital ecosystems \cite{Swan2015, Zheng2018, Espinosa2022, Jahid2023}. It links blocks through cryptographic mechanisms, each one containing transaction data among network participants or nodes. These transactions are recorded and formed into new blocks, then validated by specialised nodes like miners, validators, or delegates and added to the blockchain \cite{Panda2021}. Blockchains are classified as either public (permissionless), allowing open access and participation, or private (permissioned), with restricted access \cite{Xu2023}.

Public blockchains, such as Bitcoin \cite{Pilkington2016} and Ethereum \cite{Ethereum2023}, stand out for their decentralised structure, offering transparency, security, and immutability, suitable for applications including smart contracts \cite{Wang2019}. The decentralised nature of these systems requires \emph{consensus algorithms} to maintain trust and proper network functioning. A consensus algorithm establishes rules for nodes in a distributed network to agree on the system's state. In blockchain, these algorithms are crucial for verifying and validating transaction blocks, ensuring network integrity and trust. Common consensus algorithms include Proof of Work (PoW) \cite{Dwor1992, Back2002, Nakamoto2008}, Proof of Stake (PoS) \cite{Ethereum2023pos}, and Delegated Proof of Stake (DPoS) \cite{Larimer2014dpos, Bitshares2018dpos}. The selection of an algorithm is influenced by security, scalability, energy efficiency, and desired decentralisation level in a blockchain network.

The widely recognised PoW algorithm selects miners by requiring nodes to solve complex mathematical puzzles and submit solutions swiftly. Solving these puzzles demands substantial computational power, with the node that successfully solves the puzzle being the first to gain the privilege of mining the next block. However, PoW exhibits a significant limitation, as it consistently favours nodes with the highest computational power, thus challenging the achievement of a truly decentralised and equitable system.

To address the shortcomings of PoW, the PoS consensus algorithm, as referenced in \cite{Saleh2021, Nguyen2019, Thin2018}, was introduced. PoS chooses validators based on their staked utility tokens, ensuring they do not manipulate the blockchain for personal gain. Validators confirm blocks and stake bets on them, with rewards proportional to their stakes. PoS operates on a staking-based incentive model. DPoS differs from PoS by having network users vote for delegates to validate blocks, enhancing democracy but potentially affecting decentralisation. While PoS and DPoS improve upon PoW by not requiring extensive hardware for block validation, they still face particular challenges.

One of the key limitations of PoS stems from the subjective and imprecise nature of stake values. While stake values are expressed numerically, their interpretation is influenced by human perception, resulting in inherent vagueness and uncertainty. For instance, if a group of individuals were surveyed about their perception of a monetary amount, whether in cryptocurrencies or traditional forms, they would provide a range of responses such as `very low,' `low,' `moderate,' `high,' or `very high.' This diversity of responses underscores the intrinsic uncertainty and vagueness within human perception. Additionally, PoS confronts the challenge of nodes with higher stake values exerting undue control over the blockchain, echoing the centralisation issues of PoW. This power imbalance stifles the growth and participation of smaller stakeholders within the network.

Regardless of the limitations facing each consensus algorithm, there is one common challenge: diversification to choose miners, validators or delegates. Diversification is a crucial characteristic in the blockchain environment because it helps prevent the centralisation of power within the network. When a small group of entities controls the majority of mining or validation power, they can potentially manipulate the network for their own gain, violating the principles of decentralisation and trustlessness that are key to blockchain technology. From a security standpoint, a diverse set of miners, validators, or delegates enhances the network's resilience to attacks. By diversifying the selection of miners or validators, consensus algorithms make it more challenging for attackers to amass enough influence to execute attacks, such as the Sybil attack \cite{CHEN20221}, successfully. Furthermore, a diverse set of participants in the validation process brings different perspectives and transparency, leading to a more dynamic and resilient ecosystem.

In response to these identified limitations, we propose an innovative blockchain protocol named Fuzzychain. Fuzzychain introduces a novel concept by incorporating Fuzzy Sets (FSs) theory to represent stake values, thereby introducing a degree of fuzziness into stake-based consensus mechanisms. This pioneering approach aims to enhance the equity and security of blockchain networks. By incorporating FSs, Fuzzychain offers validators with diverse stake amounts more opportunities for periodic selection, fostering a \textbf{more inclusive and equitable system}. This unique contribution aims to mitigate the challenges associated with the imprecise nature of stake values in traditional PoS consensus algorithms, ultimately promoting a more robust and participatory blockchain network. The \textbf{highlights of this paper} are the following:
\begin{itemize}
\item An innovative consensus algorithm for blockchain networks employing fuzzification for stake determination in a proof-of-stake framework.
\item A new technique aimed at equitable stake allocation among all validators.
\item This method surpasses contemporary leading consensus algorithms in ensuring broader stake distribution while maintaining the same functionality.
\end{itemize}

Finally, an illustrative example is presented to show the performance of the equitable consensus algorithm, as well as its advantages concerning other consensus algorithms such as PoW, PoS, and DPoS. 

The paper is structured as follows: Section 2 gives a background on blockchain and fuzzy logic. Section 3 reviews related work. Section 4 details the methodology of the proposed \emph{'Fuzzychain'} and its consensus algorithm. Section 5 analyses the security of the proposed algorithm. Section 6 discusses implementation features and key results. Section 7 examines the advantages, disadvantages, and challenges of the algorithm. The paper finishes with the conclusions and future work in Section 8.

\section{Background}
This section provides concepts focusing mainly on block- chain technology and fuzzy sets, which have been used to develop this proposed work. Section \ref{Sec: blockchain} defines blockchain, public blockchain (permissionless), and elliptic curve cryptographic.  Section \ref{Sec: fuzzylogic} describes the fuzzy sets, triangular fuzzy sets, and finally, linguistic variables. 

\subsection{Blockchain technology} \label{Sec: blockchain}
Blockchain technology is a decentralised and distributed ledger system that records and verifies transactions across multiple computers or nodes in a network \cite{Zhang2024}. A public blockchain is a type of blockchain network known as a permissionless blockchains. This kind of blockchain is open to anyone and is maintained by a decentralised network of nodes (computers), where the nodes can validate transactions and contribute to the census mechanism \citep{FERDOUS2021publicandpri}. To add a new block to the blockchain, nodes must agree on the validity of the transactions through a consensus mechanism. Various consensus mechanisms, such PoW \cite{Dwor1992, Back2002, Nakamoto2008} and PoS \cite{Ethereum2023pos}, are used to achieve this agreement. 

\begin{figure}
	\centering
		\includegraphics[scale=.35]{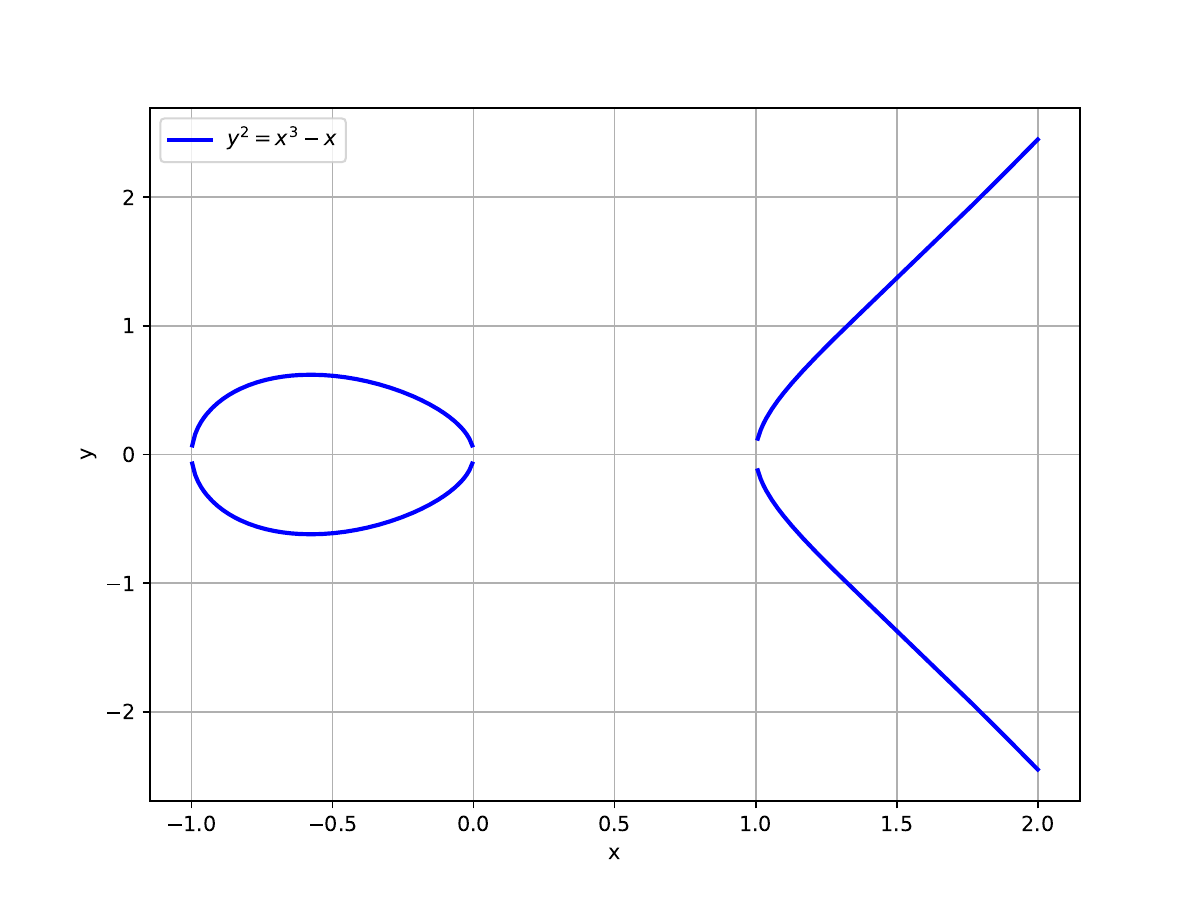}
	\caption{The figure depicts the elliptic curve defined by $y^2=x^3-x$.}
	\label{Fig: ecc}
\end{figure}

Blockchain often has high levels of security due to its decentralised nature and the use of cryptography. Cryptography plays a critical role in blockchain technology, which is applied to secure transactions, protect data, and control access to the blockchain. Specifically asymmetric cryptography is used to generate two keys: a public key and a private key, these keys are occupied to authenticate users and sign transactions, among others \cite{Menezes1996}. There are different asymmetric encrypted algorithms. The most popular are ElGamal \cite{Elgamal1985}, RSA \cite{RSA1978}, and Elliptic Curve Cryptography (ECC) \cite{Koblitz1994}.  ECC has been widely used because it uses smaller parameters but with equivalent levels of security than other algorithms, obtaining advantages such as faster
computations and smaller keys \cite{Johnson2001}. According to \cite{Koblitz1994}, an elliptic curve is defined as follows and depicted in Figure \ref{Fig: ecc}.

\begin{definition}
    Let $K$ be a field of characteristic $\neq 2,3$, and let $x+ax+b$  be a cubic polynomial with no multiple roots, where $a,b \in K$. An elliptic curve over $K$ is the set of points $(x,y)$ with $x$, which satisfy the equation: $$y^2=a^3+ax+b$$  with a single element denoted by $\odot$ and is called the point at infinity. 
\end{definition}

Blockchain offers multiple properties such as transparency and security, among others, nevertheless, still it faces challenges like scalability limitations \cite{Sanka2021} due to the high computational requirements and efficient consensus mechanisms \cite{Espinosa2022}.

\subsection{Fuzzy sets} \label{Sec: fuzzylogic}
Fuzzy sets theory, introduced by Zadeh in 1965 \cite{Zadeh1965}, extends the classical set theory to handle uncertainty and vagueness by allowing elements to have degrees of membership rather than just being either fully in or out of a set. This extension is particularly valuable in situations where precise classification is difficult due to ambiguity or imprecision in the data.

A fuzzy set is a collection of items where each element has a membership value that represents the degree to which it belongs to the set. These membership values range between 0 and 1, where 0 indicates no membership (completely outside the set) and 1 indicates full membership (completely inside the set). Values between 0 and 1 represent partial membership, indicating varying degrees of belongingness. The next paragraph provides a formal definition of a fuzzy set according to \cite{Aisbett2010, Mendel2017}.

\begin{figure}
	\centering
		\includegraphics[scale=.32]{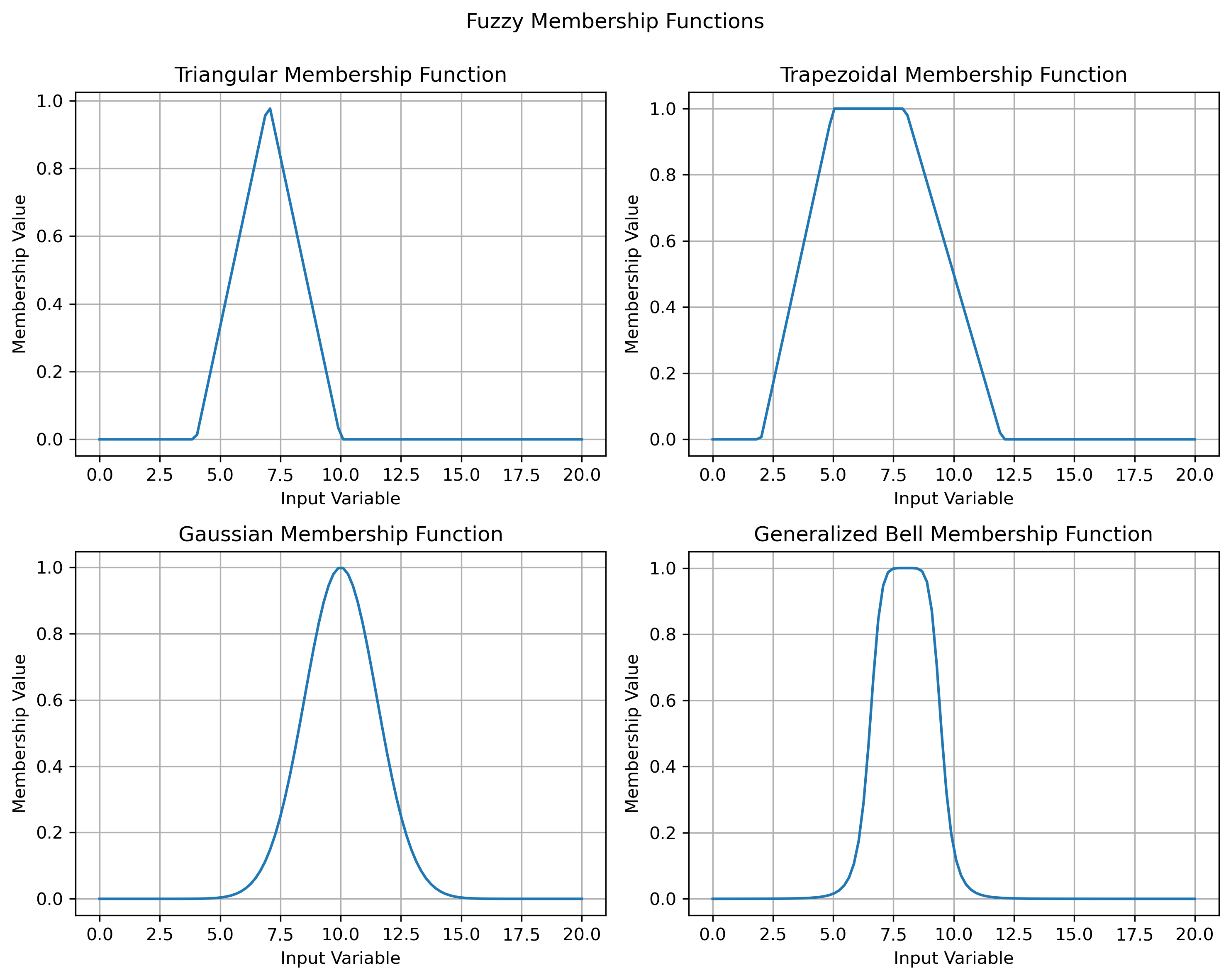}
	\caption{The figure depicts four fuzzy membership functions.}
	\label{Fig: fourmf}
\end{figure}

\begin{definition}
    Let $X$ be a universe set. $A$ is a fuzzy set if exist a function $\mu_A:X   \rightarrow [0,1]$ such that $$A=\{(x,\mu_A(x)):x \in X\}.$$

where $\mu_A$ denotes the membership function of $A$ and $\mu_A(x)$ is called the degree of membership, or membership grade, of $x$ in $A$.
\end{definition}
There are different membership functions to represent fuzzy sets, such as triangular, trapezoidal, Gaussian, and Generalised Bell membership functions, among others, as shown in Figure \ref{Fig: fourmf} \cite{Mendel2017, Jain2020}. The use of approximated assessments, such as fuzzy values, has shown that very accurate values are unnecessary \cite{Delgado1998}. Therefore, using triangular fuzzy membership functions is common and simpler. Consequently, for the development of experiments and testing the performance of the proposed consensus algorithm (see Section 6), the triangular fuzzy membership function is employed, as defined below \cite{Mendel2017}:

\[\mu_A(x)=\begin{cases}
0 & \text{ if } x<a \\ 
\frac{x-a}{b-a} & \text{ if } a\leq x\leq b \\ 
\frac{c-x}{c-b} & \text{ if } b\leq x\leq c \\ 
 0& \text{ if } x>c 
\end{cases}\]

One of the most interesting uses of fuzzy logic and fuzzy sets theory was given by Zadeh \cite{Zadeh1999} when he proposed the idea of computing with words (CWW), ``\textit{a methodology in which the objects of computation are words and propositions drawn from a natural language}''. The words in this paradigm CWW may be modelled using linguistic variables. According to \cite{Mendel2017}, a linguistic variable is a variable whose values can take words or sentences in a natural language and may be represented by fuzzy sets.

\begin{definition}
A linguistic variable is characterised by a quintuple $(L,T,X,G,\mu)$, where: 
\begin{itemize}
    \item[-] $L$ is the name of the variable,
    \item[-] $T$ is the set of linguistic terms of $L$,
    \item[-] $T$ is the universe of discourse,
    \item[-] $G$ is a syntactic rule that generates linguistic terms of $L$, and
    \item[-] $\mu$ is a semantic rule that associates each linguistic term $t \in T$ its meaning, $\mu(t)$, which is a fuzzy set on $X$.
\end{itemize}

\end{definition}

Notice that, $\mu$ can be seen as a function $\mu: T \rightarrow F(X)$, where $F(X)$ denotes the set of fuzzy sets of $X$, one fuzzy set for each $t \in T$.

\section{Related work}

Both in the domain of blockchain technology and others \cite{Liu2021}, numerous studies have underscored the pivotal role played by consensus algorithms in shaping the success and viability of blockchain. Notably, Saleh et al. \cite{Saleh2021} have accentuated the limitations inherent in PoW algorithms while advocating for PoS as a prominent alternative for achieving a more balanced and sustainable equilibrium within blockchain networks. In the context of PoS, participants are categorised as either validators or stakeholders, a distinction that fosters a more harmonious and efficient ecosystem.

One of the primary advantages of PoS over PoW is its inherent security against certain attack vectors. In PoS, users aiming to become validators must commit a portion of their stake as collateral, thereby subjecting their potential gains to risk. However, a well-recognised deficiency of PoS systems is the vulnerability to scenarios in which a malicious user, through legitimate investment or illicit means such as secret block creation \cite{Eyal2018} or selfish mining \cite{Gemeliarana2018}, amasses enough stake to control a majority (51\% or more) of the PoS blocks. This level of control not only jeopardizes the integrity of the blockchain but can also disrupt its operations, as vividly described by Larimer in \cite{Larimer2013}.

The vulnerability of PoS systems to such attacks is primarily attributed to the deterministic nature of stake values, which typically manifest as precise, unambiguous quantities consistently ranked in the same order. However, introducing a degree of fuzziness into the evaluation of stake values could address this issue. By incorporating fuzziness into the ranking of stake values, the certainty of maintaining a higher stake becomes uncertain, making it more challenging for a single user to dominate the blockchain.

Furthermore, the comparative analysis of PoW and PoS conducted in \cite{vashchuk2018} highlights the sustainability concerns associated with PoW-based blockchains. The sheer energy consumption of PoW, which in 2019 equated to the energy requirements of an entire country like Denmark, accentuates the urgency of exploring more energy-efficient alternatives. PoS, with its focus on stake rather than computational power, presents a greener and more environmentally responsible approach. However, the transition to a PoS-based model introduces its unique set of challenges, notably related to the potential competition in stake values.

In response to these formidable challenges, our innovative solution, Fuzzychain, aims to obviate the need for participants to engage in a cutthroat race for higher stake values by introducing a "halo of fuzziness" into the quantification of stake. This novel concept strives to strike a nuanced balance between stake-based consensus mechanisms and the vulnerabilities they may entail, ultimately contributing to the stability, security, and sustainability of blockchain networks. Fuzzychain represents a significant stride in the ongoing exploration of consensus algorithms, aiming to enhance the resilience and equitable operation of blockchain systems.

\section{Fuzzychain: equitable consensus algorithm} \label{Sec: methodology}
In this section, we propose the Fuzzychain: an equitable consensus algorithm. A general description is given in Subsection \ref{Sec:gd} and the specific details are further explained in Subsection \ref{Sec:sd}.

\subsection{General description} \label{Sec:gd}
\begin{figure}
	\centering
		\includegraphics[scale=.75]{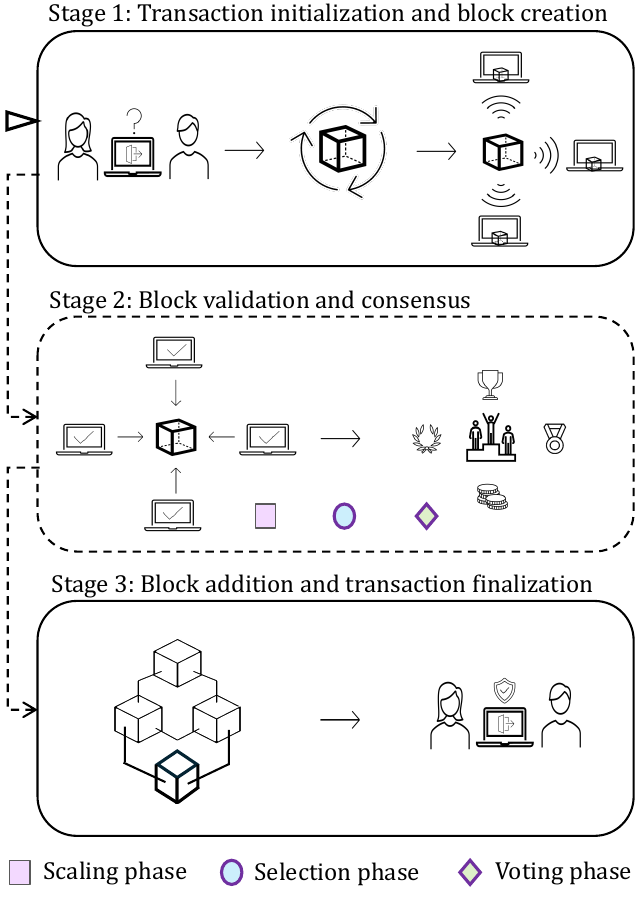}
	\caption{The figure shows the blockchain process explained in three stages. Stage 1: Initialise the transaction, generate a new block, and send it to the participating network. Stage 2: Block validation and consensus process. In this stage, three phases are executed: scaling, selection and voting.  Stage 3: The new block was added to the blockchain and transaction finalisation.}
	\label{Fig: diagrambt}
\end{figure}

In order to provide a clearer explanation of Fuzzychain, it has been segmented into three distinct stages, as depicted in Figure \ref{Fig: diagrambt}.\\

\noindent \textit{Stage 1: Transaction initialisation and block creation}

In the first stage of Fuzzychain, the process begins with the initialisation of transactions. These transactions represent digital agreements, exchanges of value, or any form of data that participants within the blockchain network wish to record. Each transaction contains details such as the sender, recipient, the amount involved, and a digital signature to ensure its integrity and authenticity. Once these transactions are collected and validated, a new block is generated. This block acts as a container, grouping a set of transactions. The creation of a new block involves cryptographic processes that secure the data within it, making it tamper-proof. After the new block is constructed, it is disseminated across the network to all participating nodes. This stage establishes the foundation for blockchain operations as it assembles the transactions and prepares them for validation.\\

\noindent \textit{Stage 2: Block validation and consensus}

The second stage of Fuzzychain is the block validation process. In this critical phase, a consensus algorithm comes into play to determine the authenticity and validity of the transactions included within the newly created block. Consensus is a fundamental concept in blockchain technology, as it ensures that all participants in the network agree on the order and content of transactions. Various consensus algorithms can be employed, such as PoW, PoS or DPoS, depending on the blockchain's design. The consensus algorithm checks the transactions for compliance with the network's rules and validates that the participants involved have the necessary permissions and resources. Once consensus is achieved, a collective decision is made on which participants (often referred to as miners, validators, or delegates) will have the responsibility to add the new block to the blockchain. This phase is essential for maintaining the integrity and security of the blockchain, preventing fraudulent or erroneous transactions from being added.\\

\noindent \textit{Stage 3: Block addition and transaction finalization}

The final stage, Stage 3, marks the process of adding the newly validated block to the blockchain, thus finalizing the transactions it contains. Once the consensus algorithm has verified the transactions and designated the responsible participants, they undertake the task of appending the new block to the existing blockchain. This process ensures that the transactions are immutably recorded sequentially and chronologically. The added transactions are considered complete, and the agreed-upon changes to the blockchain state take effect. The added block becomes a permanent part of the blockchain's history, forming a secure and transparent ledger of all network activities. The blockchain's value lies in this stage, as it guarantees the reliability and trustworthiness of the recorded transactions, enabling the blockchain network to maintain its integrity and functionality.

\subsubsection*{General Performance of Fuzzychain}
These three stages collectively form the core of the Fuzzychain protocol, providing a systematic and secure approach to handling transactions within a blockchain network. By breaking down the process into these distinct stages, Fuzzychain enhances the transparency and reliability of blockchain operations, offering a practical solution to the challenges associated with trust, stake value, and control.

This research work will focus on Stage 2, a pivotal step in the functioning of a blockchain network. During this stage, a consensus algorithm is employed to ensure the integrity and security of the network. Through this intricate process, various nodes or validators participate in verifying the transactions and aiming to validate the block. This verification process is critical as it enhances the transparency and immutability of the blockchain, ensuring that only legitimate transactions are added to the distributed ledger. Once a consensus is reached among the participating nodes, the network can collectively decide on the next valid block to be added, thereby reinforcing the decentralised nature of the blockchain ecosystem.

\begin{figure}
	\centering
		\includegraphics[scale=.65]{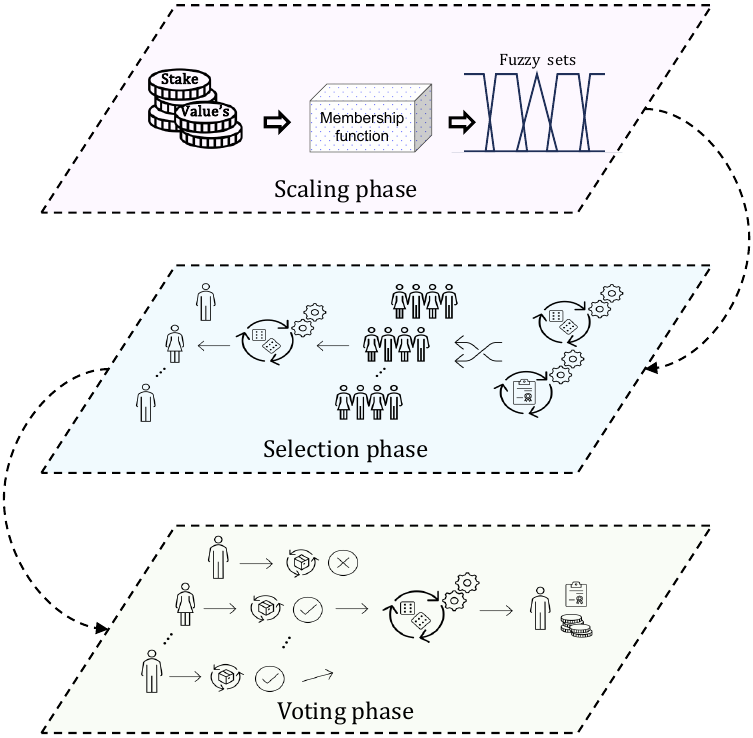}
	\caption{The figure illustrates an overview of the scaling, selection, and voting phases employed in the development of the equitable consensus algorithm. These phases are applied during the second stage, where block validation and consensus occur.}
	\label{Fig: phases}
\end{figure}

Fuzzychain employs an equitable consensus algorithm based on proof of stake to validate block transactions. In the proposed algorithm, a numerical value representing a participant's stake is scaled into a set of fuzzy sets by using a membership function. Each numerical value is then assigned to an associated linguistic label, introducing an element of fuzziness to the stake representation. Figure \ref{Fig: phases} illustrates an overview of the proposed equitable consensus algorithm. Subsequently, a set of participants is chosen based on their reputation from each fuzzy set. From each set of participants, selected randomly one or two validators, engage in the validation process and indicate whether the block should be accepted or rejected.

To determine the block's validity, a voting mechanism is employed, with consensus reached based on the majority of participants' decisions. If the majority indicates acceptance, the block is added to the Fuzzychain; otherwise, it is rejected. Subsequently, from the successful participants, one is randomly chosen to add the new block to the blockchain. These successful participants maintain their reputations and the selected participant who added the block receives a commission for completing the validation process. Conversely, unsuccessful participants are penalised, decreasing their reputation. Consequently, in the next round, they will be less likely to be chosen. A detailed description is provided in the following section for a more comprehensive understanding of the proposed equitable consensus algorithm.

\subsection{Equitable consensus algorithm; design and specifications} \label{Sec:sd}
Our proposal aims at introducing an equitable consensus algorithm based on the proof of stake and fuzzy sets theory to validate and verify block transactions, thereby giving rise to the first Fuzzychain. Such an equitable consensus algorithm is composed by three phases: scaling, selection and voting phase, depicted in Figure \ref{Fig: phases} and further detailed below.

\subsubsection*{Scaling phase}
In this phase, the procedure entails scaling a participant's stake, represented by the numerical value into a set of fuzzy sets through the application of a membership function (MF) (in our case a triangular fuzzy membership). Therefore, it is defined the linguistic variable used by the consensus algorithm.
\begin{definition}\label{Def: fsd}
    Let $L$ be the linguistic variable defined by the Participant's stakes. The set of linguistic terms of $L$ is $T=\{T_1, T_2, T_3, \dots, T_n\}$ and the universe of discourse is the interval $X=\left[ l,r\right]$ with $l<r$ and $r,l \in \mathbb{N}$.
\end{definition}

For the linguistic variable L, one example of the set of linguistic terms, T, could be $T=\{Very\: Low, Low, $ $Moderate, High, Very\: High\}$. The terms in $T$, for instance, \textit{Very low}, \textit{Low}, \textit{Moderate}, etc. can be called linguistic labels (LL).  

The first step of this phase is to divide the universe of discourse $X$ into $n$ uniformly spaced and distributed type-1 fuzzy membership functions (T1-MFs), where each T1-MF is related with a corresponding fuzzy set. In the next step, any user's stake's numeric value $x$ is scaled and located in a fuzzy set defined on $X$, and each stake value is identified by a linguistic label. Since the value $x$ may belong to different fuzzy sets with different degrees of membership, the following definition is presented.

\begin{definition}\label{Def: mdh}
    \textbf{Highest membership degree function }. Let $x$ be a stake value and let $T_1, T_2, T_3, \dots, T_n$ be fuzzy sets defined on the scale $l$ to $r$. The highest membership degree function (HMDF) of the element $x$ across these fuzzy sets is defined as:
    $$HMDF(x)=\max\{\mu_{T_1}(x),\mu_{T_2}(x),\mu_{T_3}(x),\dots,\mu_{T_n}(x)\}$$
    where
    \begin{itemize}
        \item[-] $\mu_{T_i}(x)$ represents the degree of membership of the element $x$ in fuzzy set $T_i$.
    \end{itemize}
\end{definition}

\begin{figure}
	\centering
		\includegraphics[scale=.62]{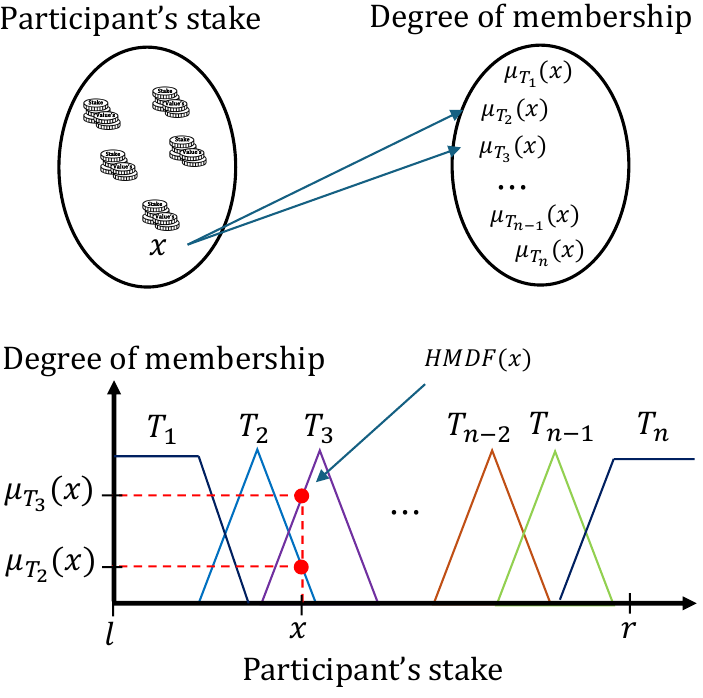}
	\caption{The figure displays the scaling phase in the equitable consensus algorithm used in Stage 2. The items $x$ in the set of Participant's stakes are scaled through the fuzzy membership function $\mu_{T_i}(x)$ into fuzzy sets $T_i$ on the universe of discourse defined from $l$ to $r$ for the linguistic variable "Participant's stake". The degree of membership for $x$ is the highest membership degree function (HMDF)}
	\label{Fig: diaalgorithma}
\end{figure}

According to Definition \ref{Def: mdh}, the numeric value $x$ of the participant's stake corresponds to a unique fuzzy set assigned through the highest membership degree function. Furthermore, each fuzzy set is identified with a unique linguistic label.  At this moment,  the items $x's$ in the set of participant's stake values have been scaled into fuzzy sets $T_i$, each bearing the appropriate linguistic label (Very low, Low, Moderate, among others), using the adequate fuzzy membership function $\mu_{T_i}(x)$. The scaling phase can be visualised in Figure \ref{Fig: diaalgorithma} and is computed in Algorithm \ref{Algorith: scalingphase}. Continuously with the proposal, the selection phase is described as follows.

\begin{algorithm}
\small
\caption{ScalingPhase (x)}
\begin{algorithmic}[1]
	\Require   Stake value $x$ ;  
	\Ensure  Assigning $T_i$ ;
		\Function{SP}{$x$}
            \For{$j=1$ \KwTo $numberParticipant'sStake$}
            \State $DM[]$;
            \State $DMH[]$;
            \State $L=Participant's \:\: stake$ ;
		    \State $T(L)=[T_1, T_2, T_3, \dots, T_n]$ ; 
            \For{$i=1$ \KwTo $n$}
                \State $ DM \gets \mu_{T_i}(x_j) $ ;
            \EndFor
            \State $DMH \gets \max(DM);$
            \State $T_i \gets ScaleStake(x_j,DMH)$ ;
        \EndFor
            \State \textbf{return} $T_i$  ;
    	\EndFunction
\end{algorithmic}\label{Algorith: scalingphase}
\end{algorithm}

\subsubsection*{Selection phase}
This phase aims to choose participants $t_i$ from each fuzzy set $T_i$. To achieve it, the phase involves two selection algorithms
\begin{enumerate*}[label=(\roman*)]
   \item validator random selection and
   \item validator selection according to the reputation.
\end{enumerate*}
The former is an algorithm that chooses the participants randomly, and the latter is already taken to choose participants based on their reputation from each fuzzy set. Reputation is a key concept in the development of this proposed algorithm, therefore the following definition is presented.
\begin{definition}
    \textbf{Reputation range.} Let $t_i$ be a participant in the fuzzy set $T_i$. For all $t_i \in T_i$, the reputation $rep(t_i,j)$ at the round $j$, is defined on interval $[0,1]$.
\end{definition}

Each participant $t_i$ entering a specific fuzzy set starts with an initial reputation set to $1$ (maximum reputation). Subsequently, their reputation may be maintained or decreased contingent upon their performance, i.e., success or failure in their validation and verification tasks. To model the behaviour of the reputation when the validator's reputation differs from 1 the following function is defined.

\begin{definition}
    Let $\eta$ be the decrease rate and let $\frac{\eta}{l}$ $,l\in \mathbb{N}$ be the increase rate. If $rep(t_i,j)$ is the reputation for the validator $t_i$ in the round $j$, then  $rep(t_i,j+1)$ for the round $j+1$ is defined by
    \[rep(t_i,j+1)=\begin{cases}
1 & \text{if } \text{$t_i$ is a  successful validator}   \\
   & \text{and $rep(t_i,j)=1$}\\ 
 rep(t_i,j) +\frac{\eta}{l}& \text{if $t_i$ is a  successful validator}\\  
 rep(t_i,j) - \eta & \text{if $t_i$ is a  unsuccessful}\\
 & \text{validator}
\end{cases}\]
where $0 \leq rep(t_i,j+1) \leq 1. $
\end{definition}

\begin{figure}
	\centering
		\includegraphics[scale=.60]{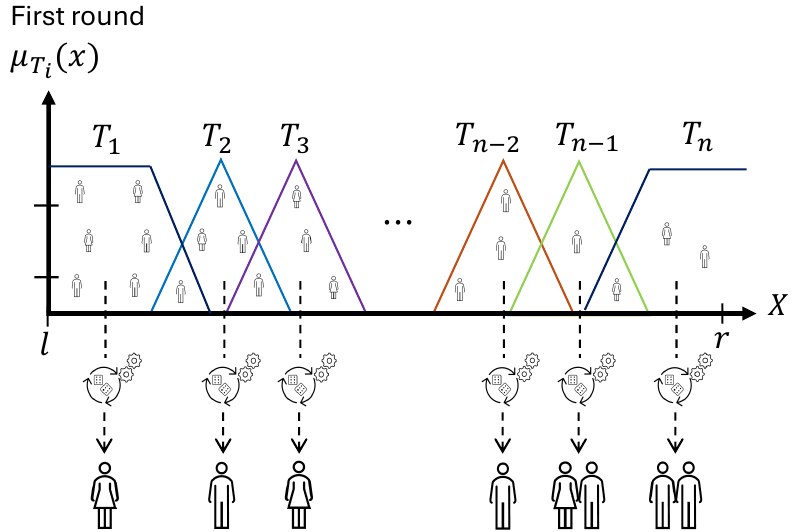}
	\caption{This figure displays the selection phase during the first round, where the validator random selection algorithm is applied to choose the participants from the fuzzy sets $T_i$, respectively.}
	\label{Fig: firstround}
\end{figure}

In order to choose the participants $t_i$, in the first round (to validate the first block), when all the participants have the same reputation, the validator random selection algorithm is applied to select one participant from each fuzzy set $T_1, T_2,\cdots, T_{n-2}$ and two participants from each fuzzy set $T_{n-1}$ and $T_n$.
Figure \ref{Fig: firstround} shows the selection process for the first round. 
\begin{rmk}
    This selection is based on the assumption that participants in fuzzy sets with the highest stake percentages have a greater interest in ensuring the network functions effectively and securely. Consequently, these participants are more trusted than those with lower stake percentages in the verification and validation process. Therefore, selecting an additional participant from the fuzzy sets with the highest stakes helps prevent participants with lower stakes from gaining control of the network, decreasing the risk of the 51\% attacks. Moreover, this selection ensures an odd number of participants,  which is crucial for the voting phase because, in the voting process is not possible to get a tie, Section \ref{Sec: securityanalysis} will explain it in more detail.  
\end{rmk}

For the next $jth$ rounds, to select the participants $t_i$ the reputation is considered, viz., the participants that have the highest reputation in each fuzzy set $T_i$ have more probability of being chosen than participants with a lower reputation. The candidates are selected using the validator selection algorithm according to their reputation, which will explain to continue. This algorithm generates two subsets $A_i$ and $B_i$ from each fuzzy set $T_i$ and both are defined as follows:
$$A_i=\{ t_i \in T_i: rep(t_i,j)=1\},$$ 
$$B_i=\{ t_i \in T_i: rep(t_i,j)\leq1\}.$$ 
Notice that the subset $A_i$ contains only participants with the highest reputation, while the subset $B_i$ includes participants with reputations less than 1 as well as those with the highest reputations, hence $A_i \subseteq B_i$. These sets are constructed with the intention that participants with the highest reputation are more likely to be chosen than those with a lower reputation.

Once the subsets are defined, the random selection algorithm chooses two participants from subset $A_i$ and one from subset $B_i$. In this way, the participants with the highest reputation have a higher probability of being chosen compared to the participants with a lower reputation. While participants with lower reputations may have fewer opportunities, but they still can excel in subsequent tasks and improve their reputations. Hence, they may ascend to the group with the highest reputation. Nevertheless, if one participant continues incorrectly doing the tasks, they will be expelled from the group of validators or even from the network. To manage this case, the validators have an error rate based on their reputation, which is presented in the following definition.

\begin{definition} \label{Def: expulsionrate}
    \textbf{Expulsion rate.} Let $rep(t_i,j)$ be the reputation of the validator $t_i$. The expulsion rate of $t_i$, $E(t_i)$, is defined by:
   \[ E(t_i)=\left\{\begin{matrix}
 0& If \:rep(t_i,j)=1 \\ 
 1-rep(t_i,j)& If \: rep(t_i,j)\neq 1 
\end{matrix}\right.\]
    
\end{definition}
As a consequence of Definition \ref{Def: expulsionrate}, the next exclusion condition is presented.

\begin{definition}
\textbf{Exclusion condition.} Let $\epsilon$ be the expulsion rate defined and allowed in the fuzzychain. If the $E(t_i)>\epsilon$ then the participant $t_i$ is excluded from the set of validators.
\end{definition}

\begin{figure}
	\centering
		\includegraphics[scale=.60]{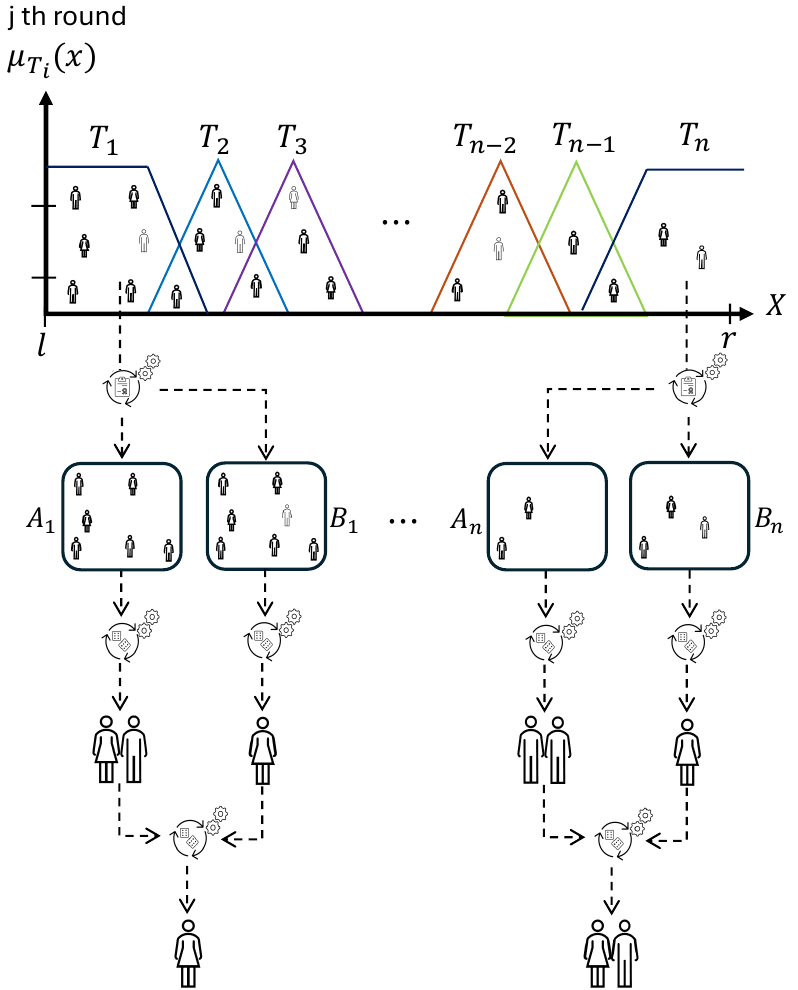}
	\caption{The figure depicts the validator selection algorithm according to the reputation to choose the participants $t_i$ from each fuzzy set $T_i$ during the j-th round.}
	\label{Fig: secondround}
\end{figure}

In the final step of this phase, participants from both subsets are combined, and the random selection algorithm is applied. It selects one participant from the sets $T_1, T_2, \cdots,$ $T_{n-2}$, two participants for $T_{n-1}$ and others two for $T_n$. Figure \ref{Fig: secondround} illustrates the process for the j-th round and is detailed in Algorithm \ref{Algorith: selectionphase}.

\begin{algorithm}
\small
\caption{SelectionPhase (x)}
\begin{algorithmic}[1]
	\Require   Fuzzy set $T_i$ ;  
	\Ensure  Validator $V_i$ ;
		\Function{SP}{$x$}
            \State $rep[]=1$ ;
            \If{$round j==1$}
                \For{$i=1$ \KwTo $n-2$}
                    \State $ t_i\gets randomlyChosenOne(T_i) $ ;
                \EndFor   
                \State $ t_{n-1}\gets randomlyChosenTwo(T_{n-1}) $;
                \State $ t_{n}\gets randomlyChosenTwo(T_{n}) $
            \Else 
                \State $A_i \gets generateSubsetA(T_i)$;
                \State $B_i \gets generateSubsetB(T_i)$
                \For{$i=1$ \KwTo $n-2$}
                \State $a_i\gets reputationChosenTwo(A_i) $ ;
                \State $b_i\gets reputationChosenOne(B_i) $ ;
                \EndFor
                \State $a_{n-1}\gets reputationChosenTwo(A_{n-1}) $ ;
                \State $b_{n-1}\gets reputationChosenOne(B_{n-1}) $ ;
                \State $a_{n}\gets reputationChosenTwo(A_{n}) $ ;
                \State $b_{n}\gets reputationChosenOne(B_{n}) $ ;
            \EndIf
            \For{$i=1$ \KwTo $n$}
                \State $M[i] \gets mix(a_i,b_i)$ 
                \State $ V_i\gets randomlyChosenOne(M) $ ;
            \EndFor
            \State \textbf{return} $Validators$ $V_i$ ;
    	\EndFunction
\end{algorithmic}\label{Algorith: selectionphase}
\end{algorithm}

 After the participants are selected for each fuzzy set, the validation and verification process of the block ensues. To do this, the voting phase is used and described in detail in the following paragraphs.

\subsubsection*{Voting phase}
The selection phase chooses one participant for the first $n-2$ fuzzy sets and two participants for the last $n-1$ and $n$ fuzzy sets. Therefore, the number of participants in the voting mechanism depends on the number of fuzzy sets, then there is an important requirement related to the fuzzy sets. The number of fuzzy sets on $X$ should be an odd number, this is crucial, particularly in the context of the voting phase within the consensus algorithm. This stipulation aligns with the design of the voting mechanism, which facilitates the selection of participants in a balanced and equitable manner during the voting phase. With an odd number of fuzzy sets, there will always be a clear majority when it comes to decision-making, minimising the likelihood of ties. In essence, the requirement for an odd number of fuzzy sets on $X$ serves to optimise the efficiency of the consensus algorithm, particularly in the critical voting phase where decisions are made regarding the acceptance or rejection of blocks within the Fuzzychain network.

Once the validators have been selected from each fuzzy set $T_i$ at the selection phase, every one of them individually engages in the validation process and subsequently indicates whether the block should be accepted or rejected. This phase involves a voting mechanism employed to determine the block's validity, wherein a consensus is reached based on the majority of participants' decisions. If the majority indicates acceptance, the block is accepted; otherwise, it is rejected. 

The sample spaces of the voting mechanism is $\Omega=\{ accepted, rejected\}$; there are no other possibilities. The validators who won the vote will be considered successful, that is if the majority indicates that the block is accepted or rejected. Therefore, to encapsulate this idea, the next definition is presented.

\begin{definition}
    \textbf{Successful validator}. A validator $V$ that participates in the voting mechanism is considered a successful validator if and only if $V$ is in the group of the validators who secure the majority vote.
\end{definition}

\begin{definition}
    \textbf{Unsuccessful validator}. A validator $V_u$ that participates in the voting mechanism is considered an unsuccessful validator if and only if $V_u$ is in the group of the validators who secure the minority vote.
\end{definition}

When the validation process is finished and it has been decided whether the block is accepted or rejected, a new selection process is carried out among the successful participants to know which of them is the winner because only one of them can take the full reward. The selection process is performed by the validator random selection algorithm presented in the selection phase.

\begin{figure}[h]
	\centering
		\includegraphics[scale=.57]{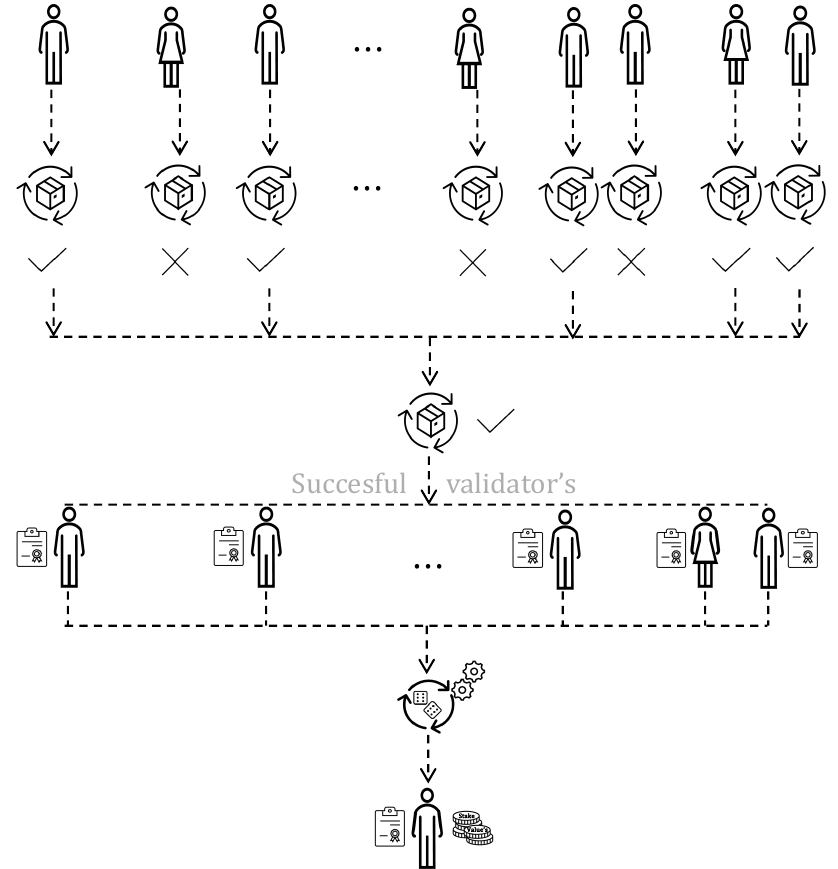}
	\caption{The figure depicts the voting phase between participants chosen in the selection phases. Each participant verifies and validates the transactions in the block and then casts their vote to accept or reject the block. Then, the winner validator is obtained by employing the random selection algorithm.}
	\label{Fig: diaalgorithmd}
\end{figure}

The reward for all successful validators is an increase in their reputation if the reputation is less than 1 and maintaining their reputation if the reputation is equal to 1. Nevertheless, for the winner validator, in addition to the reputation, a commission for having completed validation is obtained. The voting phase is depicted in Figure \ref{Fig: diaalgorithmd} and computed in Algorithm \ref{Algorith: votingphase}.  

\begin{algorithm}
\small
\caption{VotingPhase (x)}
\begin{algorithmic}[1]
	\Require   Validator $V$ ;  
	\Ensure  Validator $V_{winner}$ ;
		\Function{VP}{$x$}
            \State $listDecision[]$;
            \State $listValidators[]$; 
            \For{$i=1$ \KwTo $NumValidators$}
                    \State $listValidators[i]$ $\gets$ $Validator$ $V_i$ $validate$ $the$ $block$;
                    \State $listDecision[i]$ $\gets$ $Validator$ $V_i$ $accepted$ $or$ $rejected$;
            \EndFor  
            \State $Res \gets votingMechanism(listDecision)$;
            \If{$Res==True$}
                \State $Transactions$ $block$ $is$ $accepted$;
            \Else    
                \State $Transactions$ $block$ $is$ $rejected$;
            \EndIf
            \State $ Succ_v \gets chosenSuccessValidators(listValidators) $;
            \If{$V \:is \:Succ_v$}
                \State $increaseReputation(V)$;
            \Else    
                \State $decreaseReputation(V)$;
            \EndIf
            \State $ V_{winner}\gets randomlyChosenOne(Succ_v) $;
            \State $Reward(V_{winner})$ ;
            \State \textbf{return} $V_{wineer}$ ;
    	\EndFunction
\end{algorithmic}\label{Algorith: votingphase}
\end{algorithm}

It is important to note that in this algorithm the increase and decrease in reputation are not proportional, reputation increases more slowly and decreases more quickly. Unsuccessful validators are penalised by lowering their reputation, which limits them to having less chance of being selected the next time. In this proposal, any participant can make a mistake, so the penalty is the same for everyone regardless of whether you have more or less stake. 

In summary, the proposed equitable consensus algorithm for Fuzzychain combines elements of proof of stake and fuzzy set theory to achieve a fair and reliable method for validating block transactions. By introducing linguistic labels to represent participants' stakes and utilizing reputation as a selection criterion, the algorithm establishes a balanced approach to participant involvement in the validating process. The incorporation of a robust voting mechanism, where consensus is reached through the majority decision of selected participants, adds an additional layer of reliability to the validation process.

The algorithm ensures that successful participants not only contribute to the blockchain by adding accepted blocks but also receive dual rewards in the form of a commission for validations and an increase in reputation. Conversely, unsuccessful participant validators face penalties, including a reduction in reputation, impacting their chances of selection in subsequent rounds. This approach encourages participants to engage actively in the network, ensuring a balanced distribution of opportunities. 

The detailed description of the proposed equitable consensus algorithm provides a foundation for understanding its inner workings and sets the stage for further implementation and optimisation. This algorithm stands as a key component in Fuzzychain's pursuit of a secure, transparent, and inclusive blockchain network. In the next section, a security analysis is presented.
\section{Security Analysis} \label{Sec: securityanalysis}
Ensuring the robustness and security of the proposed equitable consensus algorithm for Fuzzychain is paramount for its successful deployment in blockchain networks. This section conducts a comprehensive security analysis to assess the algorithm's resilience to potential threats and its ability to maintain the integrity of the network.

\begin{itemize}
    \item[] \textit{Untrustworthy validators.} The security of blockchain networks relies on the assumption that a significant portion of the validators are honest and act in the best interest of the network. If a large majority of validators collude or behave maliciously, they could potentially compromise the integrity of the blockchain. To mitigate these risks,  blockchain protocols often set a threshold of honest validators required for the network to operate securely. This threshold could be expressed as a certain percentage of the total of the set of validators. Nevertheless, Fuzzychain does not focus on the total percentage of validators, on the contrary, it focuses on some fuzzy sets being reliable. In Fuzzychain the specific threshold of honest validators required to operate securely is determined by a minimum of fuzzy sets trusted.
    \begin{definition}
        Let $n$ be the number of fuzzy sets on the universe of discourse $X$, the minimum number of fuzzy sets trusted required to operate securely is determined by  $$\left \lfloor \frac{n-2}{2} \right \rfloor+1$$ 
where $\left \lfloor \right \rfloor$ denotes the floor function.
    \end{definition}
    
This stipulation holds profound significance in fortifying the consensus process against potential threats posed by malicious activities or coordinated attacks. By mandating trust in a significant majority of fuzzy sets, the algorithm provides formidable defences, shielding the system from attempts aimed at compromising its integrity or disrupting its functionality.

    \item[] \textit{Fuzzy Stake Representation}. The introduction of fuzzy sets in representing participants' stakes adds an element of uncertainty and flexibility to the algorithm. Fuzzy representations allow for a nuanced and distributed approach to stake distribution, making it challenging for an attacker to predict or control the specific stake distribution necessary to compromise the majority of validator positions.
    \item[] \textit{Reputation-Based Selection}. The algorithm emphasises reputation as a factor in the participant selection process. Validators are chosen based not only on their stake but also on their reputation within the network. This reputation-based selection introduces an additional layer of complexity for potential attackers, as they would need to influence both stake and reputation to control the majority of validator slots. Another point of view is that the algorithm employs a mechanism that involves the randomised selection of validators from diverse groups. This decentralisation in the validator selection process prevents a single entity from gaining control over the majority of validator slots in a deterministic manner. As a result, even if an entity has a significant stake, the randomness in validator selection mitigates the risk of concentration and control.
    \item[] \textit{Incentive Structure}. The dual rewards system, combining both commissions for successful validations and reputation increases, incentivises active and honest participation. Attackers attempting a 51\% attack would risk reputational damage and reduced chances of future selection, discouraging malicious behaviour. In addition, the algorithm's randomised selection and the inclusion of reputation as a factor ensure a dynamic and ever-changing participant landscape. This dynamic nature makes it challenging for an attacker to consistently maintain control over the majority of the network's computational power.
    \item[] \textit{Continuous Improvement and Adaptability}. The penalties imposed on unsuccessful participants, including a decrease in reputation and reduced chances of selection in the next round, contribute to an environment of continuous improvement. This adaptability further deters malicious actors as the network adjusts to minimise the impact of unfavourable behaviours.
\end{itemize}

The proposed equitable consensus algorithm's resistance against 51\% attacks is rooted in its decentralised, reputation-based, and dynamic participant selection process, coupled with the introduction of fuzzy sets for stake representation. These features collectively enhance the algorithm's robustness and make it inherently challenging for any single entity to gain control over the majority of the network's computational power in a predictable or sustained manner.

\section{Experiments and Results} \label{Sec: results}
This section performs and discusses a set of experiments designed to assess the performance of the equitable consensus algorithm and the Fuzzychain proposed within a permissionless scenario. An overview of the information system, the outcomes and the findings from these experiments are presented.

The experiments were developed on the following software and computer specifications. It includes a CPU and an Intel® Core™ i7-7500U processor, featuring a clock speed of 2.70GHz and four cores. The operating system used is Ubuntu 22.04.3. For compiling, A C++ compiler, GCC version 7.4.0 is utilised. The programming language employed is Python, specifically version 3.10.12. Additionally, the system makes use of two libraries: ECDSA and SIMPFUL.

\subsection{Experimental results} \label{Sec: er}
This proposal encompasses the execution of two experiments. Experiment 1 is dedicated exclusively to displaying the performance of the equitable consensus algorithm. Experiment 2 is designed to showcase a comparison concerning the equitable consensus algorithm proposed with other consensus algorithms such as PoW, PoS, and DPoS.

\subsubsection*{Experiment 1}
The objective of this experiment is to demonstrate the performance of the equitable consensus algorithm. To accomplish this, simulations have been developed to illustrate how the consensus algorithm operates when selecting participants in each round to validate the block. Furthermore, the experiment offers insights into the frequency with which each winning participant is chosen. This analysis allows us to gain a comprehensive understanding of not only the algorithm's functionality but also the distribution of selection among participants of the different fuzzy sets, a critical element of an equitable consensus algorithm.

For this experiment, we considered 990 validators participating in the equitable consensus algorithm, with each validator assigned a stake value. According to Definition \ref{Def: fsd}, we defined the information representation scale for the stake values with bounds $l=0$ and $r=10$. We then segmented this scale into five linguistic terms: Very Low (VL), Low (L), Moderate (M), High (H), and Very High (VH).

\begin{rmk}
    For this experiment we have used distributed symmetric linguistic labels, nevertheless, the proposed algorithm algorithm allows to use of unbalanced linguistic labels \cite{Herrera2008}. 
\end{rmk}

\begin{figure}[h]
	\centering
		\includegraphics[scale=.373]{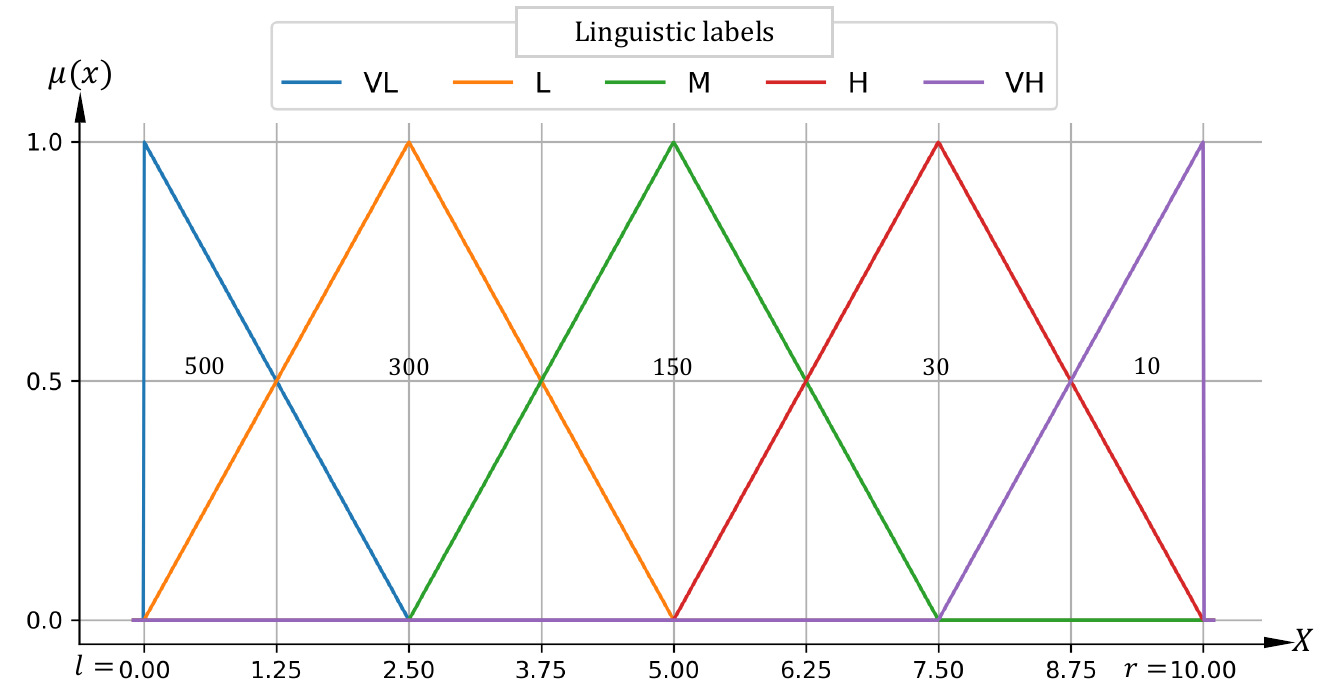}
  \caption{Membership functions for linguistic terms T(Participant's stakes)= \{Very Low (VL), Low (L), Moderate (M), High (H), Very High (VH)\}.}
	\label{Fig: linterms}
\end{figure}

Following the scaling phase in Section \ref{Sec:sd}, to develop this experiment, the validators are distributed using a triangular membership function as follows: 500 in the linguistic term `VL', 300 in `L', 150 in `M', 30 in `H' and 10 in `VH' as illustrated in Figure \ref{Fig: linterms}.

From each one of these linguistic terms, we select a set of validators based on their reputation. Nevertheless, in the first round, since all the validators have the same initial reputation set to $1$, the selection process is random. According to the selection phase presented in Section \ref{Sec:sd}, to validate the first block, one validator is chosen randomly from each linguistic variable VL, L, and M, and two validators from each linguistic variable H and V. From this set of validators, only one validator is further selected to perform the validation process. The seven validators corresponding to each linguistic term initiate the validation process, with each of them casting their vote. 

According to the voting phase in Section \ref{Sec:sd}, the block is either accepted or rejected, and the successful and unsuccessful validators are then announced. On the one hand, each successful validator receives a reputation increase of 0.005 as a reward for doing well; on the other hand, unsuccessful validators decrease their reputation by 0.1 every time they want to damage the network. Finally, from the pool of successful validators, only one is randomly chosen as the winner, who is entitled to receive a commission in addition to the increase in reputation.

\begin{figure*}[]
	\centering
	\includegraphics[scale=.43]{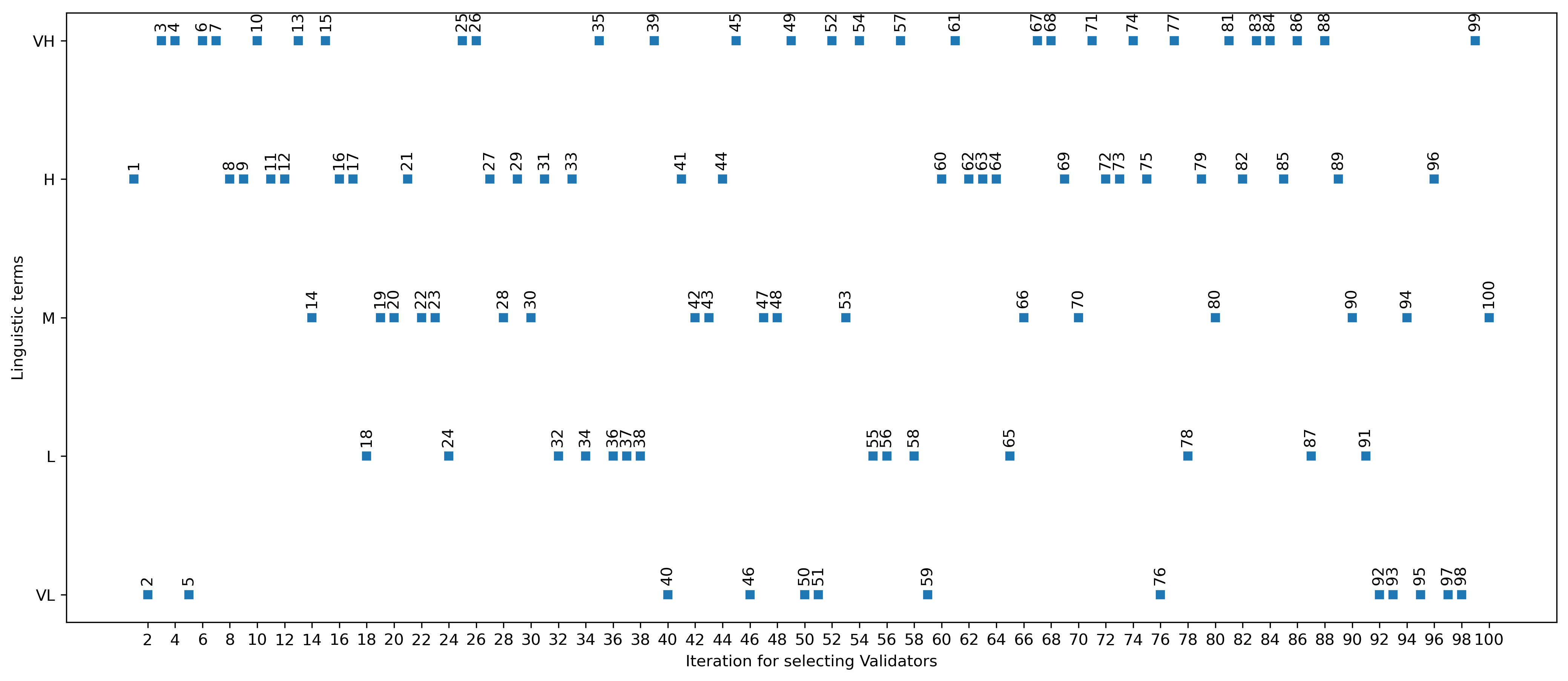}
	\caption{The figure shows the selection of validators in each iteration after running the algorithm at 100 rounds.}
	\label{Fig: selecvalidatorsiter}
\end{figure*}

\begin{figure}[]
	\centering
	\includegraphics[scale=.43]{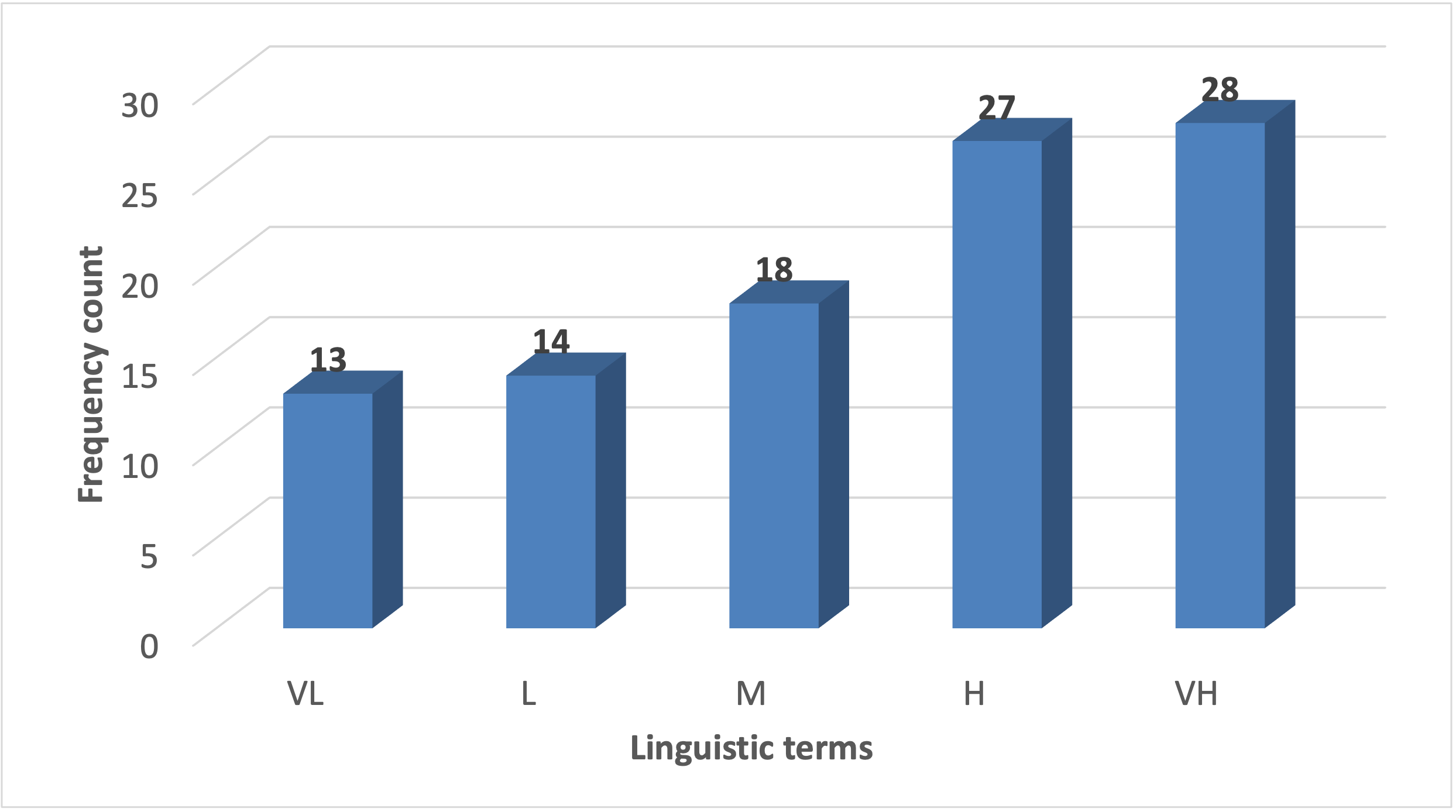}
	\caption{The figure summarises the number of validators selected for each linguistic term at 100 rounds.}
	\label{Fig: frequencyrel}
\end{figure}

For the next round, to choose the validators from each linguistic term, everyone's reputation is considered. We select a set of validators using the validator selection algorithm according to the reputation shown in the selection phase in Section \ref{Sec:sd}. Analogous to the first round, from this set of validators, only one validator is selected to perform the validation process. The seven validators corresponding to each linguistic term initiate the validation process, and each of them casts their vote. According to the voting phase in Section \ref{Sec:sd}, the block is either accepted or rejected, and both successful and unsuccessful validators are announced. Successful validators receive an increase in reputation of 0.005, while unsuccessful validators have their reputation decreased by 0.1. From the group of successful participants, one winner is randomly selected and entitled to receive a commission.

This algorithm is executed every time a block needs validation. For this experiment, the algorithm was run for 100, 200, 300, 400, and 500 rounds, aiming to demonstrate the frequency count of validators selected for each linguistic term. The validators selected change in each round, that is because in each round the reputation is recalculated and the validator with the highest reputation is chosen according to the algorithms presented in \ref{Sec:sd}. First, the outcomes of the experiment for 100 rounds are presented. Figure \ref{Fig: selecvalidatorsiter} shows the selections of validators in each iteration after running the algorithm at 100 rounds. From the figure, the proposed method shows the variations in the selection of validators. Figure \ref{Fig: frequencyrel} summarises the number of validators selected for each linguistic term at 100 rounds. The X-axis represents the linguistic terms of each validator, and the Y-axis illustrates the frequency count. From the figure, it is evident that most validators are selected from the linguistic terms H and VH. That means a greater number of validators with a higher interest in the network working well have participated in the verification and validation process. Always consider the participation of the other validators corresponding to the linguistic terms VL, L, and M. The mean of selected validators is $20$ with a standard deviation of $7.07$. Therefore, the dispersion of selection among validators for the linguistic terms VL, L, and VH are within the first standard deviation and the terms M and H are within the second standard deviation.

\begin{figure}[h]
	\centering
	\includegraphics[scale=.485]{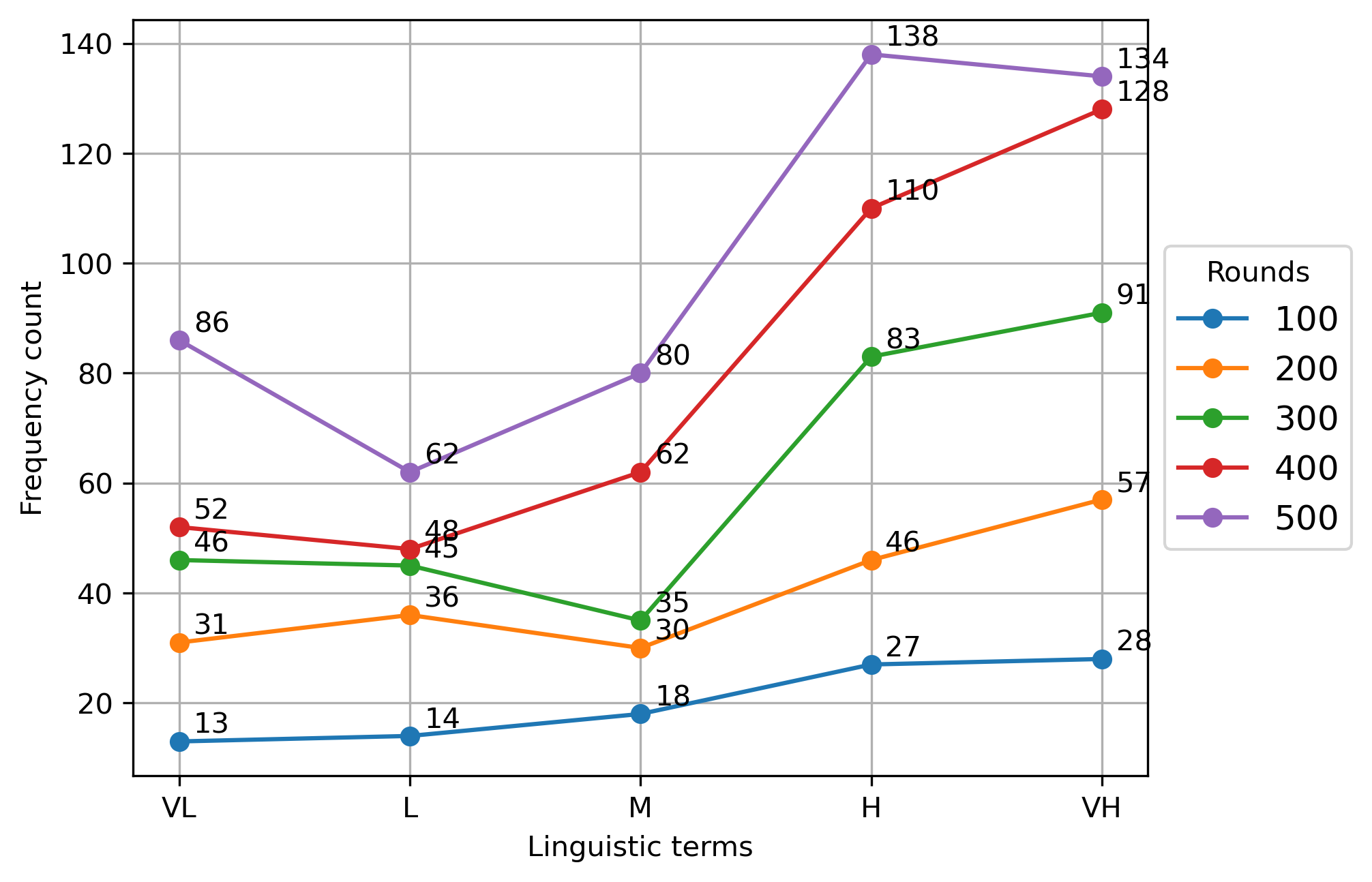}
	\caption{The figure despite five plots corresponds to the selection of validators for each linguistic label in 100, 200, 300, 400 and 500 rounds.}
	\label{Fig: selecvalidators}
\end{figure}

For rounds 200, 300, 400, and 500, the outcomes are summarised in Figure \ref{Fig: selecvalidators}. This figure, despite five plots, corresponds to the selection of validators for each linguistic label in different rounds. For instance, the green graph illustrates the number of validators selected from each linguistic term when the algorithm runs in a set of 300 rounds. In this scenario, 46 validations were done by the validators selected from the linguistic term VL, 45 from L, 35 from M, 83 from H and 91 from VH. At the different rounds, the linguistic terms `H' and `VH' participate more than others in the validation and verification process. This is beneficial as the algorithm was designed to select validators with higher reputations, higher stakes and, consequently, higher trust, thus increasing trust in the system.

\begin{figure}[h]
	\centering
	\includegraphics[scale=.37]{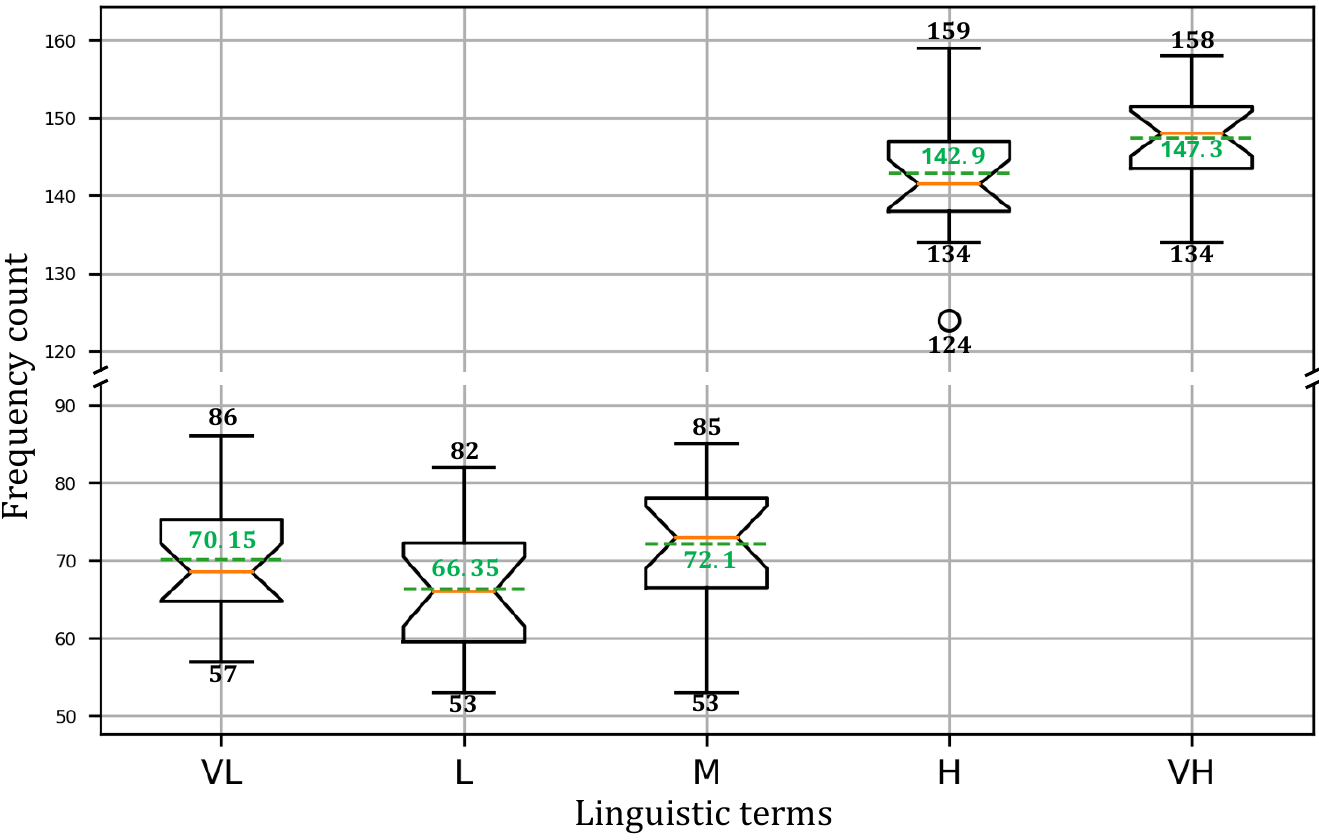}
	\caption{Figure shows a boxplot representing the frequency and distribution of the validator selected by the equitable consensus algorithm in twenty repetitions.}
	\label{Fig: boxplot}
\end{figure}

Intending to study the behaviour of the proposed algorithm, we decided to repeat 20 times the experiment where the validators are selected during 500 rounds. The idea is to show the outcomes obtained in each round, nevertheless, since the multiple numeric values, we decided to use a boxplot to group the results.  Figure \ref{Fig: boxplot} presents a boxplot for each linguistic term VL, L, M, H, VH and the Y-axis show the frequency count. From the figure, it is clear that the data in the boxplot for H and VH are greater than the data in the boxplot for VL, L, and M. For instance, for the linguistic term VH the minimum value of selected validators during the 20 repetitions is 134 and the maximum is 158 with a mean of 147.3 which is displayed with a green dashed line. For the linguistic term M, the minimum value obtained in the twenty repetitions is 53 and the maximum value is 85, where the mean is 72.1. It is important to mention this event never happens in the PoS consensus algorithm because in PoS the distribution of the validator selection process is always oriented towards validators with a higher stake. 

\subsubsection*{Experiment 2}
This section presents a comparison concerning the proposed consensus algorithm and other consensus algorithms used in blockchain. In this comparative analysis, we delve into the distinctive features of the proposed consensus algorithm with well-established blockchain consensus mechanisms, such as PoW, PoS, and DPoS. Each of these consensus algorithms operates on distinct principles and exhibits unique strengths and weaknesses. Firstly, PoW, the pioneer in consensus mechanisms, relies on computational power to validate transactions through complex mathematical puzzles \cite{Bach2018, Wendl2023}. On the other hand, PoS introduces a more energy-efficient approach, where validators are chosen based on the amount of stake they hold and are willing to "stake" as collateral \cite{Saad2021, Fahim2023}. Meanwhile, DPoS further optimises scalability by employing a selected group of delegates to validate transactions, nevertheless, this specific group of delegates can contribute to the risk of centralisation \cite{Saad2021, Alrowaily2023}. Table \ref{Tab:comparison} summarises the properties of each consensus mechanism.  This section aims to highlight the differences, enabling a comprehensive understanding of how the proposed consensus algorithm contributes to time complexity, energy consumption, security, and decentralisation and improvement of blockchain technology.

In order to provide a quantitative comparison, the consensus algorithms PoW, PoS and DPoS were simulated and set to the following conditions. For the PoW, 100 miners were selected to participate in the verification and validation process during 100 rounds.  The consensus PoW chooses the first miner who solves the mathematical proof. The frequency of selected miners was registered, and then the data were used to compute the following metrics: Gini coefficient, Skewness, and Kurtosis. These metrics also are computed for consensus PoS, DPoS, and Fuzzychain.
For the consensus algorithm PoS, a set of 100 validators is defined to participate in the validation process. In this implementation, the validators have different stakes, and the validator with a higher stake has more probability of choosing the winner to do the validation process. This experiment was executed 100 rounds and the frequency of the selected validators was registered, and then calculate the three metrics as well. 
For the consensus algorithm DPoS, 100 delegates were defined to participate in the verification and validation process. In this experiment, the validator is chosen according to the stake and the reputation; that is, the delegate that has a higher stake with a higher reputation will have a higher probability of being the winner delegated. Similar to the others, the number of rounds in this experiment was 100, consequently the frequency of selected delegates was calculated and then the metrics were computed also.  
In the case of Fuzzychain, we considered 990 validators participating in the equitable consensus algorithm, with each validator assigned a stake value. The validators are distributed using a triangular membership function as follows: 500 in the linguistic term `VL', 300 in `L', 150 in `M', 30 in `H' and 10 in `VH' as illustrated in Figure \ref{Fig: linterms}. The results are displayed in Table \ref{Tab:comparison1} where a numerical comparison is presented.

\begin{table*}[]
\begin{tabular}{lllll}
\toprule
\textbf{}                   & \multicolumn{1}{c}{\textbf{PoW}} & \multicolumn{1}{c}{\textbf{PoS}} & \multicolumn{1}{c}{\textbf{DPoS}} & \multicolumn{1}{c}{\textbf{Fuzzychain}} \\ \midrule
\textbf{Time complexity}    & High                             & Lower than DPoS                   & Lower than PoW                    & Lower than PoS                          \\
\textbf{Energy consumption} & High                             & Lower than PoW                   & Lower than PoW                    & Lower than PoW                          \\
\textbf{Security}           &      High                            &      Lower than PoW                             &                    High               &             High                            \\
\textbf{Decentralisation}   &         Lower than Fuzzychain                        &        Lower than Fuzzychain                          &                      Lower than Fuzzychain              &     High        \\        \bottomrule                  
\end{tabular}
\caption{This table summarises a qualitative comparison between PoW, PoS, DPoS, and Fuzzychain.}
\label{Tab:comparison}
\end{table*}

\begin{table}
\begin{tabular}{@{}ccccc@{}}
\toprule
                 & PoW      & PoS       & DPoS      & Fuzzychain \\ \midrule
Gini coefficient & 0.5992   & 0.4934    & 0.4126    & \textbf{0.1720}        \\
Skewness         & 1.8855  & 1.5243  & 0.9630   & \textbf{0.2243}       \\
Kurtosis         & 3.3206  & 2.8247   & 1.3253   & \textbf{-1.7489}       \\\bottomrule
\end{tabular}
\caption{This table shows the numerical comparison between PoW, PoS, DPoS, and the proposed consensus algorithm. In bold are the top scores, indicating the most favourable interpretation of these statistics in relation to equality.}
\label{Tab:comparison1}
\end{table}

Table \ref{Tab:comparison1} presents a comparison between PoW, PoS, DPoS, and the proposed consensus algorithm under Fuzzychain. To do this, the Gini coefficient is calculated for every consensus algorithm discussed before. The Gini coefficient assesses the disparity within the values of a frequency distribution, such as income levels. A Gini coefficient of 0 denotes total equality, reflecting a situation where all individuals have the same income or wealth. On the other hand, a Gini coefficient of 1 denotes maximal inequality, where all income or wealth is concentrated with a single individual, leaving none for others. Skewness near zero suggests a more symmetric and, thus, more evenly structured distribution. Additionally, the lower the kurtosis, the lesser the chances of encountering extreme values. 

In Table \ref{Tab:comparison1}, the Gini coefficient in Fuzzychain stands out as being notably lower than that of the other algorithms, indicating a higher degree of fairness in the selection of validators. The equitable consensus algorithm employed by Fuzzychain excels in promoting a more balanced distribution of validation responsibilities among network participants compared to its counterparts. A lower Gini coefficient suggests that Fuzzychain is successful in mitigating the concentration or centralization of validation power, thereby fostering a more inclusive and democratic blockchain network. This enhanced equity in validator selection is crucial for maintaining the decentralization and security of the network, as it reduces the risk of a single entity gaining disproportionate influence. The findings underscore the effectiveness of Fuzzychain's approach to consensus, emphasizing its commitment to creating a robust and fair blockchain ecosystem.

\section{Discussion}
The proposed equitable consensus algorithm for Fuzzychain presents a distinctive approach to achieving consensus in blockchain networks, blending elements of proof of stake and fuzzy set theory. One notable feature is the incorporation of linguistic labels to represent participants' stakes within fuzzy sets. This introduces a level of fuzziness, enhancing the flexibility and expressiveness of stake representation. The algorithm's emphasis on reputation as a factor in participant selection during the mining process is noteworthy, promoting a fair and inclusive approach. The randomised selection of participants from each fuzzy set adds an element of unpredictability, preventing any single participant or group from consistently dominating the validation process. The utilisation of a voting mechanism based on the majority decision of participants ensures a collective and democratic approach to block validation. The rewarding of successful participants with both a commission for validations and an increase in reputation creates a positive incentive structure, motivating active and responsible participation. Conversely, the penalties imposed on unsuccessful participants, including a decrease in reputation and reduced chances of selection in the next round, contribute to maintaining a balance and encouraging continuous improvement. 

The algorithm presents a consensus mechanism that addresses issues of fairness, security, and participant engagement in the Fuzzychain blockchain network. However, the practical implications and potential challenges of implementing such a system will require further exploration and empirical testing in real-world blockchain scenarios.

In this work, we have presented an equitable consensus algorithm. Nevertheless, it can be seen as a cryptography scheme because it can work with the different types of membership functions that exist in fuzzy sets, for instance, triangular MF, trapezoidal MF, Gaussian MF, and generalised bell MF, among others. Similarly, it is possible to change the randomly chosen algorithm to another algorithm with a better performance in choosing the participants. Even more, the voting mechanism can be modified to make a decision efficient.

Usually, each fuzzy set has a different number of participants and is probably that this number is bigger in linguistic labels such as VL, L, M than H, and VH. Another important advantage of this proposed algorithm is that the participants in the Fuzzychain may move to other fuzzy sets where the participants are less than others. This is possible because the validator receives a commission to make the process correctly. 

In the present iteration of our fair consensus algorithm, threshold values for reputation are defined as crisp, non-fuzzy values. In future works, explore the 'computing with words' technique to enhance this. The idea is to develop a system that dynamically modifies the fuzzified reputation each round based on a range of factors. These include the number of validators associated with each linguistic term, fluctuations in reputation metrics, and the average value of this data. By adopting this approach, the algorithm's resilience would be bolstered, making it more adaptable to shifts in real-world scenarios, such as sudden changes in the number of validators.

\section{Conclusion}

In this study, we have introduced an innovative approach to the PoS consensus algorithm within the blockchain domain, where validators' stakes are modelled using fuzzy logic values. The application of this methodology to a PoS blockchain simulation has yielded consistent results, as detailed in Section \ref{Sec: results}. Notably, the Fuzzychain algorithm consistently opts for validators from diverse groups, with a predominant selection from the Medium and High categories. However, unlike other PoS methods, it ensures that any group, including pourer lower stake validators, is overlooked. This deliberate approach facilitates a more extensive distribution of rewarded stakes, particularly for well-minted transactions within the Fuzzychain framework. Notably, the flexibility of the Fuzzychain allows a selected validator to belong to any group without necessitating a predetermined precise probability for group selection. This dynamic provision of opportunities ensures an inclusive and non-predetermined approach to validator selection, enhancing the overall integrity of the transaction-solving process and contributing to heightened security. Future iterations of the Fuzzychain will explore extensions involving sets capable of handling increased uncertainty, and alternative characteristics for the fuzzy stake will be considered to guide the selection process. These ongoing developments aim to enhance further the Fuzzychain's robustness and applicability within diverse blockchain-operative contexts.

\section*{CRediT authorship contribution statement}
\textbf{Bruno Ramos Cruz:} Conceptualization, Methodology, Investigation, Writing - Original Draft. \textbf{Javier Andreu-Perez:} Conceptualization, Methodology, Writing - Review \& Editing, Supervision. \textbf{Francisco J. Quesada:} Conceptualization, Methodology, Writing - Review \& Editing. \textbf{Luis Martínez:} Conceptualization, Methodology, Writing - Review \& Editing, Supervision.

\section*{Declaration of competing interest}

The authors declare that they have no known competing financial interests or personal relationships that could have appeared to influence the work reported in this paper.

\section*{Data availability}
Data will be made available on request.

\section*{Acknowledgement} 
We express our sincere gratitude to Dr Prashant K. Gupta and Dr Deepak Sharma for the early discussions about this work with Dr Javier Andreu-Perez.

\bibliographystyle{model1-num-names}
\bibliography{cas-refs}

\end{document}